\documentclass[twocolumn,showpacs,showkeys,preprintnumbers,amsmath,amssymb,aps,prb]{revtex4}

\usepackage{graphicx}
\usepackage{epsfig}

\begin{document}

\title{Magnetic and electronic properties of double-perovskites and
estimation of their Curie temperatures by \textit{ab initio} calculations}

\author{Tapas Kumar Mandal,$^1$ Claudia Felser,$^2$ Martha
Greenblatt,$^1$  and J\"urgen K\"ubler$^3$$^*$}
\affiliation{$^1$Department of Chemistry and Chemical Biology,
Rutgers, The State University of New Jersey, 610 Taylor Road,
Piscataway, NJ 08854, USA} \affiliation{$^2$Institut f\"ur
Anorganische und Analytische Chemie, Johannes
Gutenberg-Universit\"at, D-55128 Mainz, Germany}
\affiliation{$^3$Institut f\"ur Festk\"orperphysik,Technische
Universit\"at Darmstadt, D-64289 Darmstadt, Germany}

\email{jkubler@fkp.tu-darmstadt.de}

\date{\today}

\begin{abstract}

First principles electronic structure calculations have been
carried out on ordered double-perovskites Sr$_{2}\rm B'\rm B''\rm
O_6$
 (for $\rm B'$ = Cr or Fe and $\rm B''$ 4d and 5d transition metal elements)
 with increasing number of valence electrons at the B-sites, and on Ba$_{2}\rm Mn\rm Re\rm O_6$
 as well as Ba$_{2}\rm Fe\rm Mo\rm O_6$. The Curie temperatures are estimated \textit{ab initio} from the
 electronic structures obtained with the local spin-density functional
 approximation, full-potential generalized gradient approximation
 and/or the LDA+U method (U - Hubbard parameter).
 Frozen spin-spirals are used to model the excited states needed to
 evaluate the spherical approximation for the Curie temperatures.
 In cases, where the induced moments on the oxygen was found to be large,
 the determination of the Curie temperature is improved by additional exchange
 functions between the oxygen atoms and between oxygen and $\rm B'$ and $\rm B''$ atoms.
 A pronounced systematics can be found among the experimental and/or calculated Curie
 temperatures and the total valence electrons of the transition metal elements.

\end{abstract}

\pacs{75.10.Lp,75.30.Et,71.20.Be}

\keywords{double-perovskites, half-metallic ferroimagnets,
electronic structure, Curie temperature}

\maketitle

\section{Introduction \label{p1}}

Recently, the report of room temperature colossal
magnetoresistance (CMR) in the ordered double-perovskite,
Sr$_2$FeMoO$_6$ has attracted enormous research
interest.\cite{terakura} The high degree of spin polarization in
Sr$_2$FeMoO$_6$ and other magneto resistive materials is believed
to be due to the half-metallic nature of these materials.
\cite{terakura,okimoto,park} Half-metallic materials as proposed
by de Groot \textit{et al.} \cite{degroot} are metallic with
respect to one spin direction whereas the other spin direction is
insulating/semiconducting leading to 100 percent spin polarization
at the Fermi energy ($E_F$). At the same time, K\"ubler \textit{et
al.} \cite{kubwill} also recognized nearly vanishing minority-spin
densities at the $E_F$ and suggested about peculiar transport
properties in Heusler alloys.

\begin{figure}[!]
\begin{center}
\epsfig{file=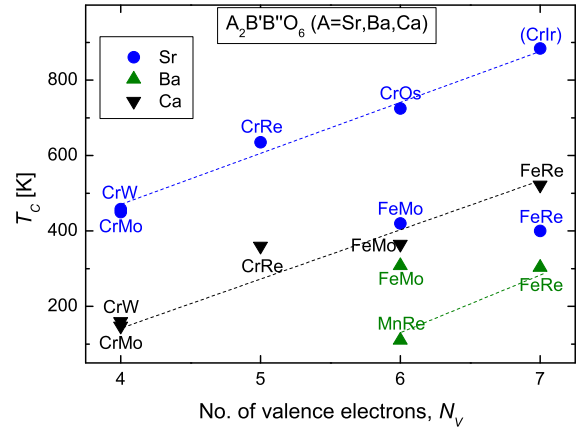,width=9cm} \caption{(Color online) Measured
Curie temperatures ($T_C$) versus the number of valence electrons.
The names of the elements $\rm B'\rm B''$ in the double-
perovskites A$_{2}\rm B'\rm B''\rm O_6$ appear near the data
points. The value in parenthesis is a projected one. }
\label{mes-Tc}
\end{center}
\end{figure}

It is known that the giant magnetoresistance GMR can be enhanced
in high spin polarized materials. Therefore, materials with a high
spin polarization are essential for applications in spintronic
devices. Among various families of materials with promising
half-metallic ferro/ferrimagnetism (HMF), however, large values of
tunnelling magneto resistance (TMR) have not been reported at room
temperature, because of their low Curie temperatures ($T_C$).
Thus, finding half-metallic ferromagnetic materials with high
$T_C$ is important in the area of spintronics. Some of the
double-perovskite oxides have attracted considerable attention due
to their very high $T_C$ (well above room temperature).
\cite{serrate} For example, Sr$_2$CrReO$_6$ is known to be a
metallic ferrimagnet with highest $T_C$ of $\approx$ 635 K.~ In
search of double-perovskites with high $T_C$, we noticed an
intriguing trend (Fig. \ref{mes-Tc}) in the measured $T_C$s with
the total number of valence electrons of the transition metals. To
emphasize the observed trends we included the Curie temperatures
for Ca$_{2}\rm B'\rm B''\rm O_6$ which, however, will not be
further considered here.

In previous work \cite{kublerxx} the electronic and magnetic
structure as well as trends in the Curie temperatures of several
Heusler compounds were studied. These are fairly simple
intermetallics for which our theoretical approach was found to
work quite well. Chemically completely different are the
double-perovskites (intermetallics versus oxides). In contrast to
the Heusler compounds the double-perovskites exhibit a larger
range of physical properties: not only half-metallic magnets but
also magnetic insulators and Mott-Hubbard insulators, for which
our theoretical approach is put to a severe test here.

Thus, in this work we have studied the electronic and magnetic
structures of several ordered double-perovskites Sr$_{2}\rm B'\rm
B''\rm O_6$ with $\rm B'$ = Cr(III) and Fe(III) while the $\rm
B''$ -site is occupied by a 4d or 5d transition metal element. We
also included the interesting pair Ba$_{2}\rm Mn\rm Re\rm O_6$ and
Ba$_{2}\rm Fe\rm Mo\rm O_6$. A systematic theoretical study of the
electronic and magnetic structures of these double-perovskites has
been carried out in view of the half-metallic and
ferro/ferrimagnetic properties in these materials. The Curie
temperatures of known members were calculated \textit{ab initio}
to compare with the experimentally measured values. In addition,
the calculations have been extended to predict the $T_C$'s of
three new members.

\section{Computational Details \label{p2}}

Three different computational procedures were applied to obtain
the electronic and magnetic structure of the double-perovskites.
The simplest is the local spin-density functional approximation
(LSDA) to density functional theory.\cite{kohn,vonbarth} The
numerical work was done using the augmented spherical wave
approximation (ASW).\cite{williams} In those cases where the
magnetic moments of a perovskite under question collapsed in the
calculation and/or the expected (see e.g. ref. \cite{terakura})
half-metallic property was not obtained, the LSDA was replaced by
a full-potential generalized gradient approximation
(GGA).\cite{perdew} Here the numerical work was done with an
extension of the ASW-method due to Kn\"opfle \textit{et
al.}\cite{knopfle} In two cases metallic antiferromagnets were
obtained in contrast to experimental information.\cite{popov} In
this case the electron correlations were assumed to be
insufficiently described by both the LSDA and the GGA. Therefore,
the semi-empirical LDA+U method \cite{anisimov} was used that is
known to provide a relatively simple and good approximation to the
problem of correlated electrons. It was incorporated into the
ASW-method by Sandratskii.\cite{sand} In the present calculations
a value for the Hubbard parameter $U$ was employed that is in line
with other experience.\cite{anisimov,fang}

\begin{table*}
\caption{\label{t1} Collection of pertinent experimental and
          calculated data for 12 double-perovskites, Sr$_{2}\rm B'\rm B''\rm O_6$ and for Ba$_{2}\rm Mn\rm Re\rm O_6$
          as well as Ba$_{2}\rm Fe\rm Mo\rm O_6$. The quantity $N_V$ is
          the total number of valence electrons supplied by $\rm B'$ and $\rm B''$. The space group symmetry
          is given in the column, symm., where known. The type of the calculated magnetic
          structure is given in the column, type, where HMF stands
          for half-metallic ferrimagnet, MIN for magnetic
          insulator, and MHI for Mott-Hubbard insulator.
          The total calculated magnetic moment,
          $M_{tot}^{\rm calc}$ is given in $\mu_{\rm B}$ per formula unit. The local moment of $\rm B'$ is denoted by
          $\mathcal{L}_{\rm B'}$, that of $\rm B''$ by $\mathcal{L}_{\rm B''}$,
          and the induced moments of the 6
          O atoms by 6$\mathcal{L}_{\rm O}$, all in units of $\mu_{\rm B}$. The
          calculated Curie temperatures, $T_{C}^{\rm calc}$, and the
          experimental ones, $T_{C}^{\rm exp}$, are given in K. }

\begin{center}
\begin{tabular}{lccccccccccc}
\hline \hline
Compound &$N_V$&symm.& $a$[\AA]&type &$M_{tot}^{\rm calc}$&$\mathcal{L}_{\rm B'}$&$\mathcal{L}_{\rm B''}$  %
& $6\mathcal{L}_{\rm O}$&$T_{C}^{\rm calc}$&${T_{C}^{\rm exp}}^{(b)}$\\
\hline \\

Sr$_2$CrMoO$_6$ & 4 & $Fm\overline{3}m^{(a)}$ & 7.840$^{(a)}$ &HMF$^{(1)}$& 2.0 & 2.251 & -0.357 & 0.106 & 379 & 420 \\
Sr$_2$CrWO$_6$ & 4 & $Fm\overline{3}m^{(b)}$  & 7.832$^{(b)}$ &HMF$^{(2)}$& 2.0 & 2.557 & -0.439 & 0.082 & 434 & 458 \\
Sr$_2$CrReO$_6$ & 5 & $I4/m^{(c)}$  & 7.814$^{(c,d)}$ &HMF$^{(2)}$& 1.0 & 2.423 & -1.272 & -0.151 & 742 & 620 \\
Sr$_2$CrRuO$_6$ & 6 & $Fm\overline{3}m^{(e)}$ & 7.903$^{(e)}$ &MIN$^{(1)}$& 0.0 & 2.222 & -1.718 & -0.504 & 577$^{(4)}$ & \\
Sr$_2$CrOsO$_6$ & 6 &$Fm\overline{3}m^{(f)}$& 7.824$^{(f)}$ &MIN$^{(2)}$& 0.0 & 2.443 & -1.893 & -0.550 & 881 & 725 \\
Sr$_2$CrIrO$_6$ & 7 & $Fm\overline{3}m^{(e)}$ & 7.881$^{(e)}$ &HMF$^{(2)}$& 1.0 & 2.296 & -0.953  & -0.343 & 884 &  \\
\\
Sr$_2$FeMoO$_6$ & 6 & $I4/m^{(b)}$  & 7.894$^{(b,d)}$ &HMF$^{(1)}$& 4.0 & 3.792 & -0.355 & 0.563 & 360 & 420 \\
Sr$_2$FeTcO$_6$ & 7 &$Fm\overline{3}m^{(e)}$ & 7.919$^{(e)}$ &HMF$^{(1)}$& 2.99 & 3.665 & -1.087 & 0.415 & 160 & \\
Sr$_2$FeReO$_6$ & 7 & $I4/m^{(b)}$  & 7.877$^{(b,d)}$ &HMF$^{(1)}$& 3.0 & 3.763 & -1.028 & 0.265 & 384 & 400 \\
Sr$_2$FeRuO$_6$ & 8 &$Fm\overline{3}m^{(e)}$ & 7.919$^{(e)}$&MHI$^{(3)}$& 0.0 & 3.613 & ~1.912$^{(5)}$ & 0.02 &  &  \\
\\
Ba$_2$FeMoO$_6$ & 6 &$Fm\overline{3}mm^{(b)}$ & 8.012m$^{(b)}$ &HMF$^{(1)}$& 4.0 & 3.564 & -0.215 & 0.651 & 331$^{(4)}$ & 308 \\
Ba$_2$MnReO$_6$ & 6 &$Fm\overline{3}mm^{(b)}$ & 8.186m$^{(b)}$ &HMF$^{(3)}$& 4.0 & 4.234 & -0.622 & 0.388 & 71$^{(3)}$ & 110 \\
\\
\hline
\end{tabular}
\end{center}
\begin{center}
\begin{tabular}{lp{12cm}}
 \footnotesize $^{(1)}$From LSDA calculation.
\footnotesize $^{(2)}$From GGA calculation.
\footnotesize $^{(3)}$From LDA+U calculation.\\
\footnotesize $^{(4)}$Obtained with 6 exchange functions.
\footnotesize $^{(5)}$Orthogonal to $\rm B'$-moment.\\
\footnotesize $^{(a)}$From Arulraj \textit{et al.}\cite{arul}~
\footnotesize $^{(b)}$From Serrate \textit{et al.}\cite{serrate}~
\footnotesize $^{(c)}$From Kato \textit{et al.}\cite{kato}\\
\footnotesize$^{(d)}$ Calculated from experimental atomic volume.~
\footnotesize $^{(e)}$Assumed values. \footnotesize$^{(f)}$From
Krockenberger \textit{et
 al.}\cite{krocken}\\

\\
\end{tabular}
\end{center}
\end{table*}

The LSDA, the GGA, and LDA+U method, besides serving to describe
the electronic structure, are also employed for obtaining
low-lying excited states from the total energies of frozen
spin-spirals with wave vectors \textbf{q}, called spiral
energies.\cite{herring, sandratskii1, sandratskii}~ In this
context the validity of the adiabatic approximation is tacitly
assumed. The spiral energies are expressed as exchange energies or
exchange functions, $j_{\tau \tau'}(\mathbf{q})$, which depend on
two basis vectors of the constituent magnetic atoms, $ \tau$ and
$\tau'$, and on wave vectors that span the irreducible part of the
Brillouin zone (BZ). The quantities $j_{\tau \tau'}(\mathbf{q})$
are extracted from the spiral energies by an algorithm given in
ref.\cite{kublerx}~ For the numerical work the force theorem
\cite{heine, mackintosh}~is used, i.e. to obtain the spiral
energies the band energies are summed up to the Fermi energy for a
given spin configuration using for these calculations the
self-consistent ground-state potential.

To estimate the Curie temperatures of the double-perovskites the
spherical approximation was used, as described by Moriya
\cite{moriya} for itinerant-electron magnetism in elementary
systems, and extended to magnetic compounds by one of these
authors.\cite{kublerx} The Curie temperature in this approximation
is given by

\begin{equation}
k_{\rm B}T_{c}=\frac{2}{3}\sum\limits_{\tau}\mathcal{L}_{\tau}^{2}
\left[
\frac{1}{N}\sum\limits_{\mathbf{q}n}\frac{1}{j_{n}(\mathbf{q})}\right]
^{-1} \ . \label{equ13}
\end{equation}

\noindent Here the  exchange functions $j_{n}(\mathbf{q})$ are
eigenvalues of a secular equation and are given in the simplest
approximation in terms of three quantities sufficient for the
description of the magnetism of double-perovskites, A$_{2}\rm
B'\rm B''\rm O_6$: $j_{11}(\mathbf{q})$ describing the exchange
interaction between the magnetic $\rm B'$-atoms,
$j_{22}(\mathbf{q})$ between the $\rm B''$-atoms and
$j_{12}(\mathbf{q})$ between the $\rm B'$- and $\rm B''$- atoms.
The quantity $\mathcal{L}_{\tau}$ in equation~(\ref{equ13}) is the
local moment of the atom at site $\tau$. In principle it must be
determined self-consistently, see equation (12) in,\cite{kublerx}
but an acceptable approximation for the double-perovskites is
$\mathcal{L}_{\tau}=M_\tau$, where $M_\tau$ is the zero
temperature moment of the atom at site $\tau$. This is so, because
the magnetic moments are well localized in spite of their
originating from itinerant electrons. To support this statement
one looks at the change of the local magnetic moments when the
tilt of the nearest-neighbor moments is large; in our case the
change is less than 0.2 \% for the magnetically dominating $\rm
B'$-atoms and round about 1.6 \% or less for the weaker moments of
the $\rm B''$-atoms.

In some of the double-perovskites the induced moments on the
oxygen were found to be relatively large, nearly reaching 0.1
$\mu_{\rm B}$ per oxygen, thus adding up to a sizable amount in
the unit cell. In these cases the determination of the Curie
temperature is improved by using in addition to
$j_{11}(\mathbf{q})$, $j_{22}(\mathbf{q})$, and
$j_{12}(\mathbf{q})$ three more exchange functions, namely
$j_{33}(\mathbf{q})$, $j_{13}(\mathbf{q})$, and
$j_{23}(\mathbf{q})$ that describe the interaction between the
O-atoms, between the O-atoms and atom $\rm B'$, and the O-atoms
and atom $\rm B''$, respectively. These parameters are also
extracted from the spiral energies in a somewhat cumbersome
procedure. The details are omitted here.

\section{Results of the Calculations \label{p3}}

In presenting the results of our calculations we use as a
"coordinate" the number of total valence electrons, $N_V$, in the
double-perovskite, A$_{2}\rm B'\rm B''\rm O_6$, that are supplied
by the transition metal elements $\rm B'$ and $\rm B''$. The
remaining electrons from the A- and O-atoms remain the same in the
series under investigation, which leads us to suppress them in the
presentation. However, we will notice that chemical valence will
not obey such a simple scheme.

\begin{figure}[!]
\begin{center}
\epsfig{file=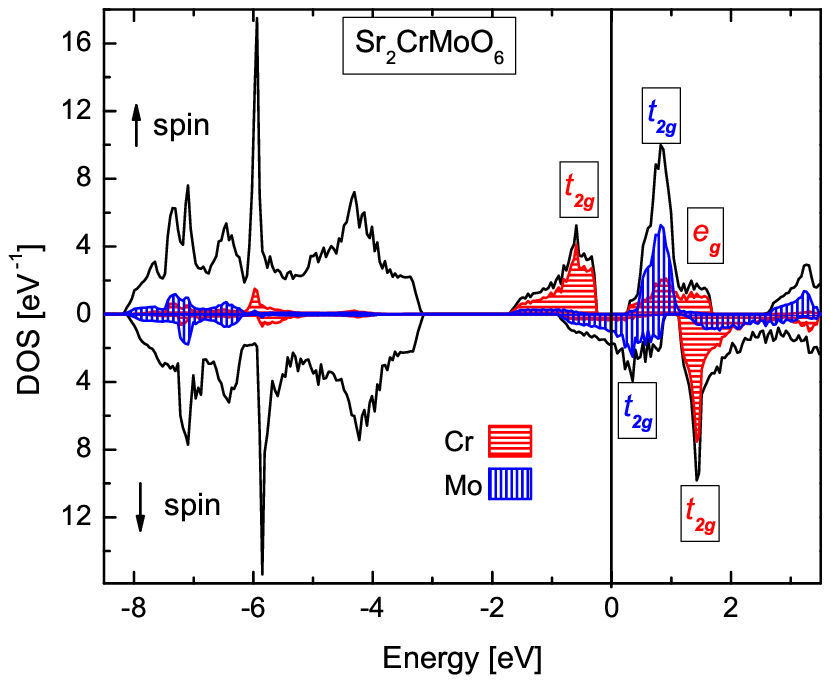,width=8cm}
\epsfig{file=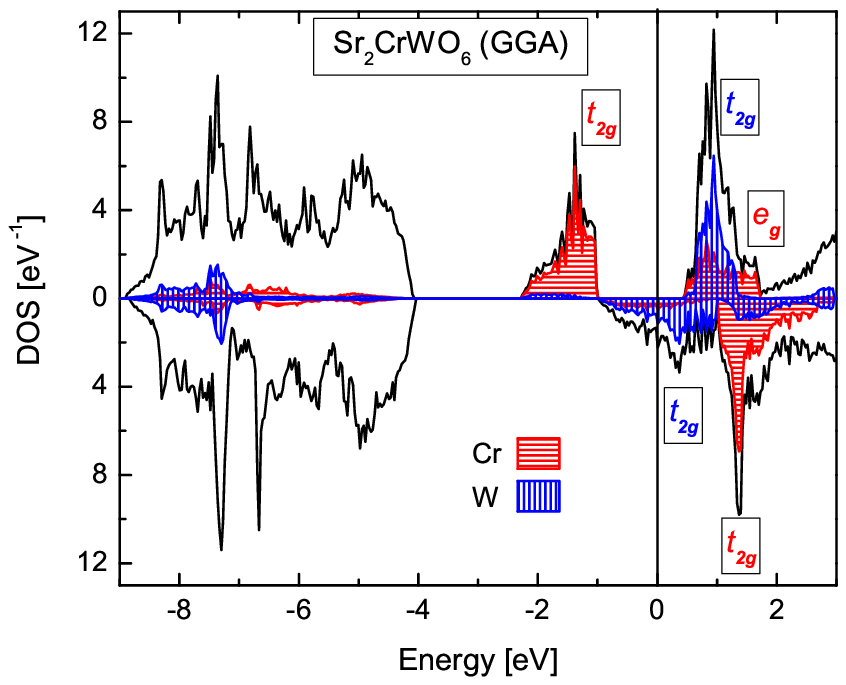,width=8cm} \caption{(Color online) Total
(black) and partial (colored) density of states of Sr$_2$CrMoO$_6$
and Sr$_2$CrWO$_6$. The total number of valence electrons from
Cr(III) and Mo(V) is $N_V$ = 4, same for Cr and W. The total
magnetic moment is in each case $M_{\rm tot}=2.0$ $\mu_{\rm B}$
per formula unit (f.u.). Symmetry labels of the parent states are
included.} \label{Dos-N12}
\end{center}
\end{figure}

\begin{figure}[!]
\begin{center}
\epsfig{file=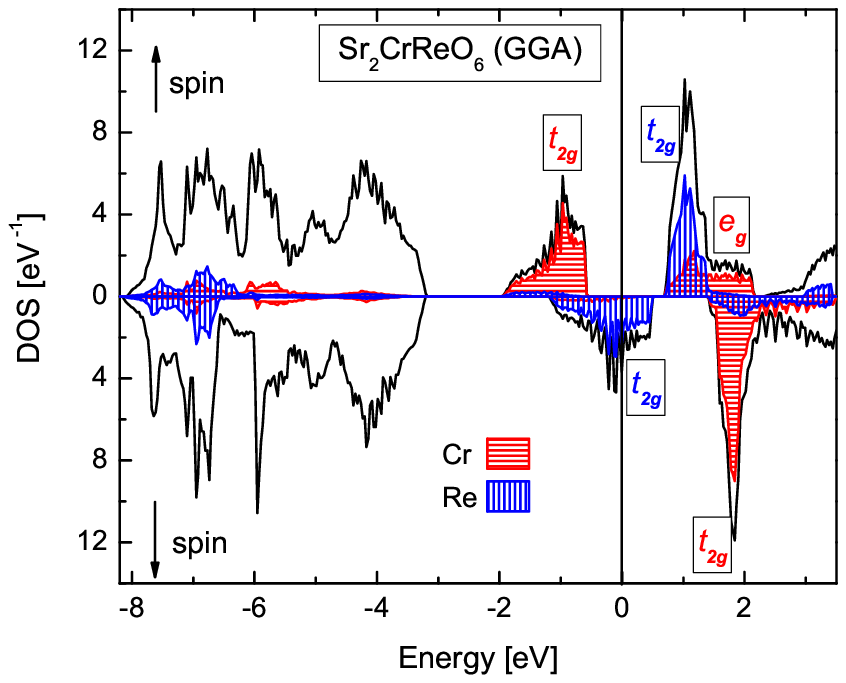,width=8cm} \caption{(Color online)Total
(black) and partial (colored) density of states of
Sr$_2$CrReO$_6$. The number of valence electrons from Cr(III) and
Re(V) is $N_V$ = 5. The total magnetic moment is $M_{\rm tot} =
1.0$ $\mu_{\rm B}$ per f.u. Symmetry labels of the parent states
are included.} \label{Dos-N13}
\end{center}
\end{figure}

\begin{figure}[!]
\begin{center}
\epsfig{file=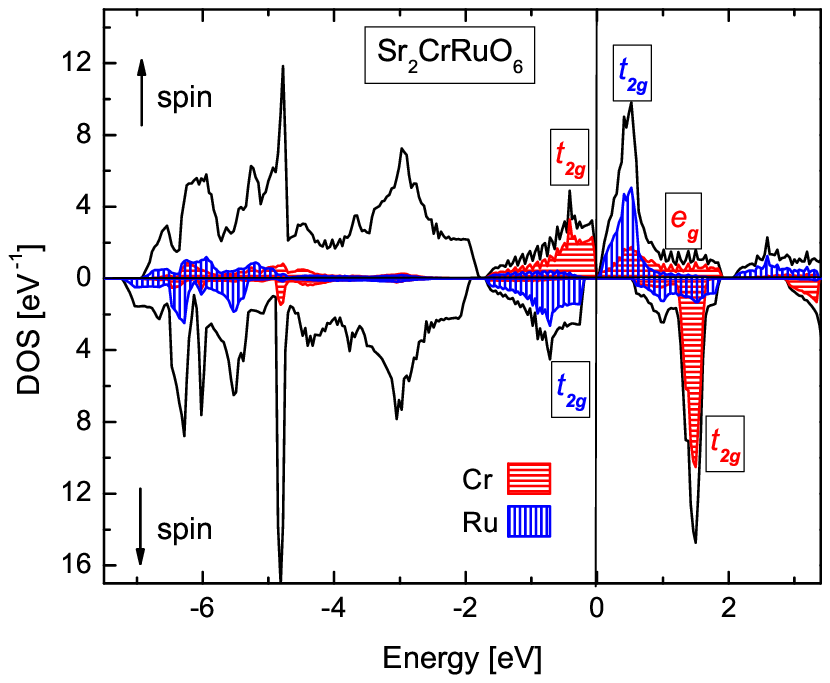,width=8cm}
\epsfig{file=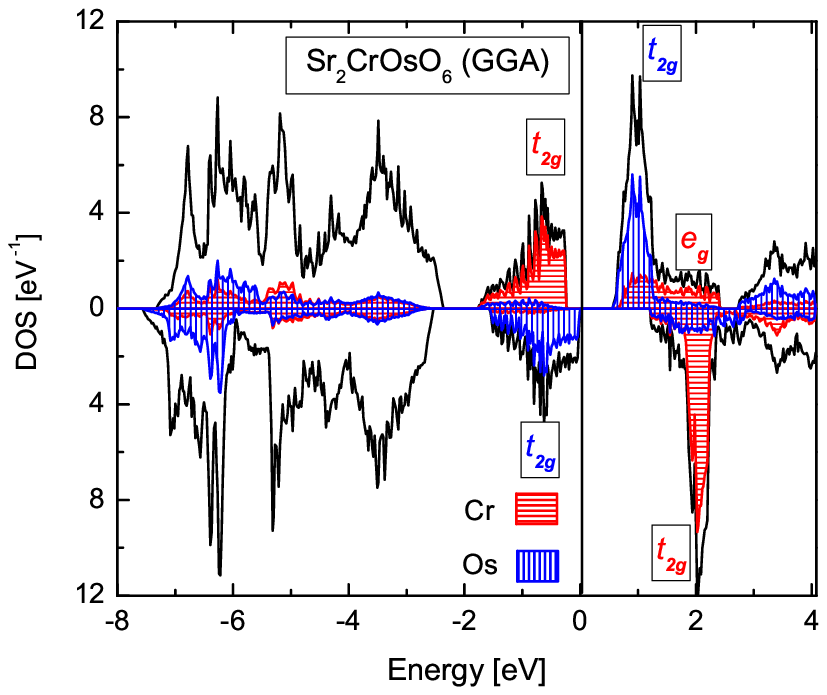,width=8cm} \caption{(Color online) Total
(black) and partial (colored) density of states of Sr$_2$CrRuO$_6$
and Sr$_2$CrOsO$_6$. The number of valence electrons from Cr(III)
and Ru(V) is $N_V$ = 6, same for Cr and Os. The total magnetic
moment in each case compensates to  $M_{\rm tot} = 0.0$ $\mu_{\rm
B}$ per f.u. Symmetry labels of the parent states are included.}
\label{Dos-N14a}
\end{center}
\end{figure}

We begin with Table \ref{t1} and the Figs. \ref{Dos-N12},
\ref{Dos-N13} and \ref{Dos-N14a}. The measured crystal structures
and lattice constants given in Table \ref{t1} were used in the
calculations except when they are tetragonal ($I4/m$). In these
cases the crystal structure was approximated by a cubic cell
($Fm\overline{3}m$) using lattice constants that were calculated
from the measured atomic volumes. It is seen that for the first
five compounds, Sr$_{2}\rm Cr\rm B'' \rm O_6$ with $\rm B''$ = Mo,
W, Re, Ru, and Os, $N_V$ increases ($N_{V} = 4$, 5 and 6), whereas
the calculated total magnetic moments, $M_{tot}^{\rm calc}$
decrease. This is so, because the individual moments of Cr and
$\rm B''$
couple antiparallel until they are entirely compensated for the compounds %
Sr$_2$CrRuO$_6$ and Sr$_2$CrOsO$_6$. The latter could thus be
called zero-moment ferrimagnets. The densities of states (DOS)
shown in Fig. \ref{Dos-N12} and \ref{Dos-N13} (Sr$_2$CrMoO$_6$,
Sr$_2$CrWO$_6$ and Sr$_2$CrReO$_6$) show a gap at the Fermi energy
in the up-spin direction (upper parts of the graphs) typical for
half-metallic ferrimagnetism (HMF). In all the DOS figures the
energy is counted from the Fermi energy. The dominant structures
at low energies (black lines) are mainly due to the O
2$p$-electrons. For Sr$_2$CrRuO$_6$ and Sr$_2$CrOsO$_6$ shown in
Fig. \ref{Dos-N14a} the half-metallic gap is found in the
down-spin electrons. Sr$_2$CrOsO$_6$ has been made by
Krockenberger \textit{et al.},\cite{krocken} who also computed the
electronic structure, which is in agreement with Fig.
\ref{Dos-N14a} as far as the insulating property is concerned.

Also insulating with zero gap in the up-spin electrons is the
compound Sr$_2$CrRuO$_6$, shown in Fig. \ref{Dos-N14a}. In Table
\ref{t1} the label MIN is understood to mean magnetic insulator.
Clearly, it is of a different kind than for instance NiO, which is
known to be a Mott-Hubbard insulator. A recent calculation by Lee
and Pickett \cite{lee} agrees fairly well with ours. Our attempts
to prepare this phase under ambient pressure have not been
successful thus far. The Curie temperature is predicted to be 577
K in the approximation with 6 exchange functions. A larger value
of 719 K is obtained in the simpler approximation with 3 exchange
functions, where the induced magnetic moments of oxygen are
ignored. The latter value, although less reliable, seems to
conform more with the trend shown in Fig. \ref{mes-Tc} than the
former.

Our electronic structure calculation of Sr$_2$CrReO$_6$ in absence
of spin-orbit coupling resulted a total magnetic moment of 1.0
$\mu_{\rm B}$ per formula unit and a half-metallic ferrimagnetic
(HMF) ground state. However, recently it has been shown
experimentally that the system is not half-metallic, in the true
sense, due to a large orbital contribution to the magnetization
(saturation was obtained only at a much higher field, $\approx$ 20
T).\cite{michalik,teresa} The measured saturation magnetization of
1.38 $\mu_{\rm B}$ for Sr$_2$CrReO$_6$ was close to the value
theoretically predicted by Vaitheeswaran \textit{et al.}
\cite{vaitheeswaran}

\begin{figure}[!]
\begin{center}
 \epsfig{file=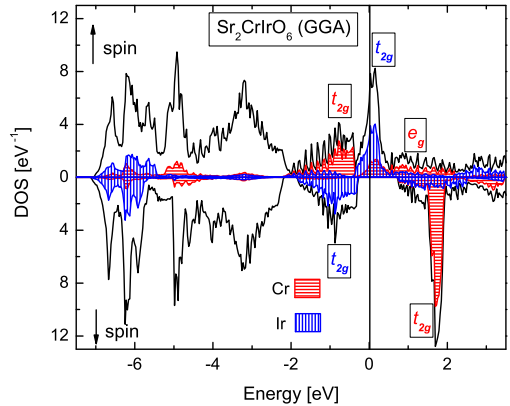,width=8cm} \caption{(Color online)
Total (black) and partial (colored) density of states of
Sr$_2$CrIrO$_6$.
 The number of valence electrons from Cr(III) and Ir(V) is $N_V$ = 7.
 The total magnetic moment is $M=1~ \mu_{\rm B}$ p.f.u.
 Symmetry labels of the parent states are included.} \label{Dos-N15b}
\end{center}
\end{figure}

The next member in the series with $N_V$ = 7, Sr$_2$CrIrO$_6$, is
an interesting case. A calculation of the electronic structure
using the GGA gives a half-metallic ferrimagnet, for which the
density of states is shown in Fig. \ref{Dos-N15b}. A calculation
of the Curie temperature results in $T_C$ = 884 K which is used as
the projected value in Fig. \ref{mes-Tc}. However, this HMF-state
is most likely metastable because a total-energy search returns a
metallic antiferromagnet of the AFII structure, which corresponds
in a k-vector notation to ${\bf{K}}=(\frac {1}{2},
\frac{1}{2},\frac{1}{2})$. In this configuration the magnetic
moments are aligned ferromagnetically in the $(111)$ plane and
alternate along the $[111]$ direction. This result has to be
questioned: correlations are likely to alter the electronic
structure and result in a Mott-Hubbard insulator. LDA+U
calculations were therefore employed but we find that even in this
case the antiferromagnet remains metallic. Comparable calculations
by Fang \textit{et al.}\cite{fang} also show that the LDA+U method
not always opens a gap at the Fermi energy.

In calculating the Curie temperature by means of equation
(\ref{equ13}) the eigenvalues $j_{n}(\bf q)$ are approximations of
the spin-wave spectrum. Negative values of $j_{n}(\bf q)$ are
unphysical and usually signal that the assumed magnetic structure
is incorrect. Since this was not seen in the present calculations
for the HMF-state, we conclude that this state is at least
metastable. Furthermore, the spin-wave spectrum for the metallic
antiferromagnet obtained by means of the LDA+U method is found to
be unphysical. Thus we discard the antiferromagnetic solution but
cannot predict that experimentally prepared Sr$_2$CrIrO$_6$ will
be a HMF.

\begin{figure}[!]
\begin{center}
\epsfig{file=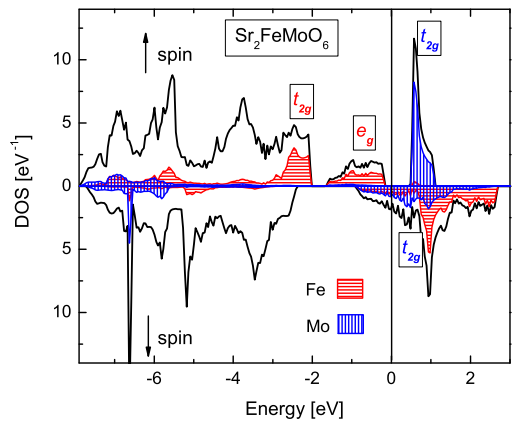,width=8cm} \caption{(Color online) Total
(black) and partial (colored) density of states of
Sr$_2$FeMoO$_6$. The number of valence electrons from Fe(III) and
Mo(V) is $N_V$ = 6. The total magnetic moment is $M_{\rm tot} =
4.0$ $\mu_{\rm B}$ per f.u. Symmetry labels of the parent states
are included.} \label{Dos-N14c}
\end{center}
\end{figure}

\begin{figure}[!]
\begin{center}
\epsfig{file=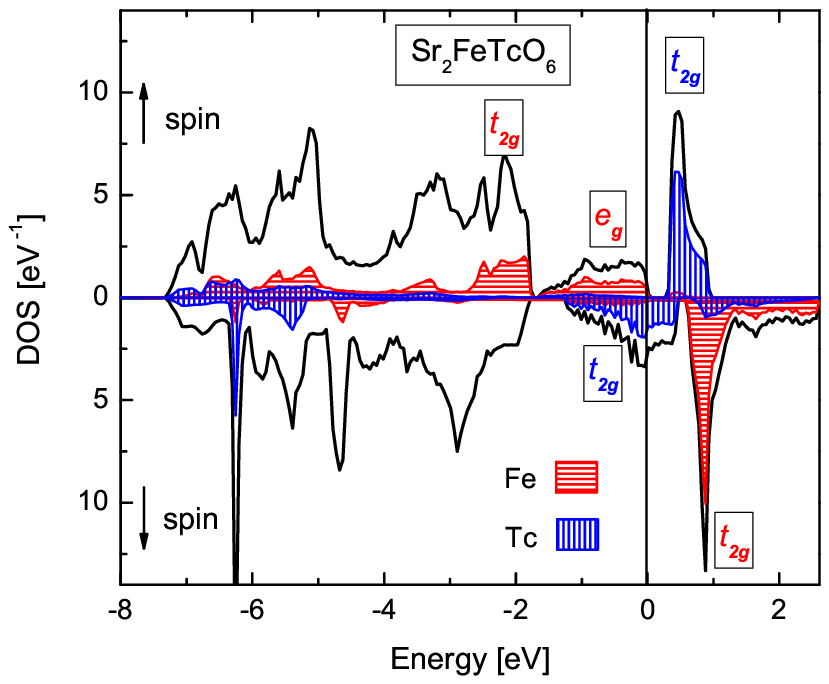,width=8cm}
\epsfig{file=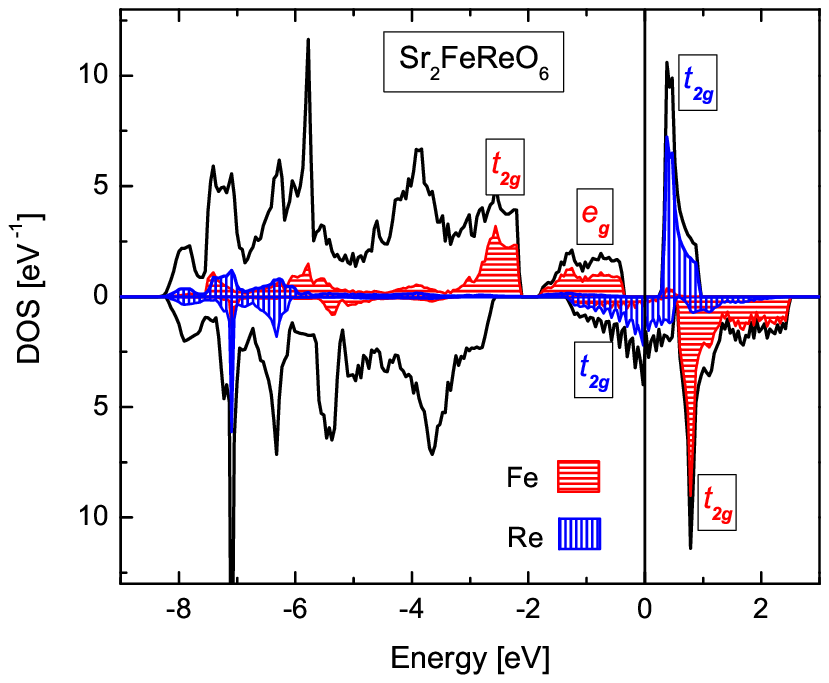,width=8cm} \caption{(Color online) Total
(black) and partial (colored) density of states of Sr$_2$FeTcO$_6$
and Sr$_2$FeReO$_6$. The number of valence electrons from Fe(III)
and Tc(V) is $N_V$ = 7, same for Fe and Re. The total magnetic
moment is $M_{\rm tot} = 3.0$ $\mu_{\rm B}$ per f.u. and nearly so
for Sr$_2$FeTcO$_6$. Symmetry labels of the parent states are
included.} \label{Dos-N15a}
\end{center}
\end{figure}

Next we concentrate on the Fe containing double-perovskites,
Sr$_{2}$Fe$\rm B''O_6$ with increasing number of valence electrons
on the transition metal atoms (Table \ref{t1}). For the  $N_V$ = 6
- double-perovskite, Sr$_2$FeMoO$_6$, we show the densities of
states in Fig. \ref{Dos-N14c}. We see HMF behavior with a large
total magnetic moment of $M_{\rm{tot}}$ = 4 $\mu_{\rm B}$ per f.u.
The DOS for Sr$_2$FeMoO$_6$ is nearly identical with that
previously obtained by Kobayashi \textit{et al.}\cite{terakura}
and Fang \textit{et al.}\cite{fang} Another earlier study by Sarma
\textit{et al.} \cite{sarma} also agrees with our calculations. In
this case, our calculation for the Curie temperature
underestimates the experimental value and gives 360 K, which is to
be compared with the experimental value of 420 K. The LSDA was
used here for simplicity, although the GGA might result in a
somewhat better estimate.

\begin{figure}[!]
\begin{center}
\epsfig{file=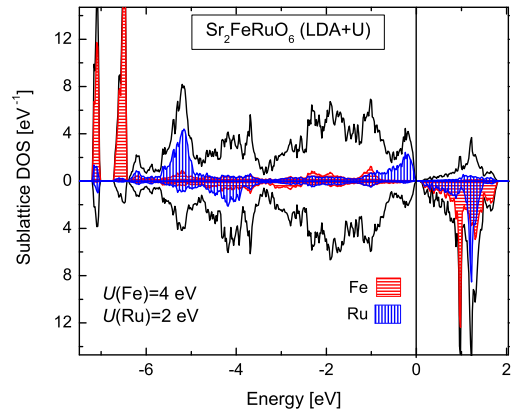,width=8cm} \caption{(Color online)
Total(black) and sublattice density of states of antiferromagnetic
(AFII or $\mathbf{K}=(0.5,0.5,0.5)$) Sr$_2$FeRuO$_6$. LDA+U was
used. The number of valence electrons from Fe(III) and Ru(V) is
$N_V$ = 8.} \label{Dos-N16}
\end{center}
\end{figure}

\begin{figure}[!]
\begin{center}
 \epsfig{file=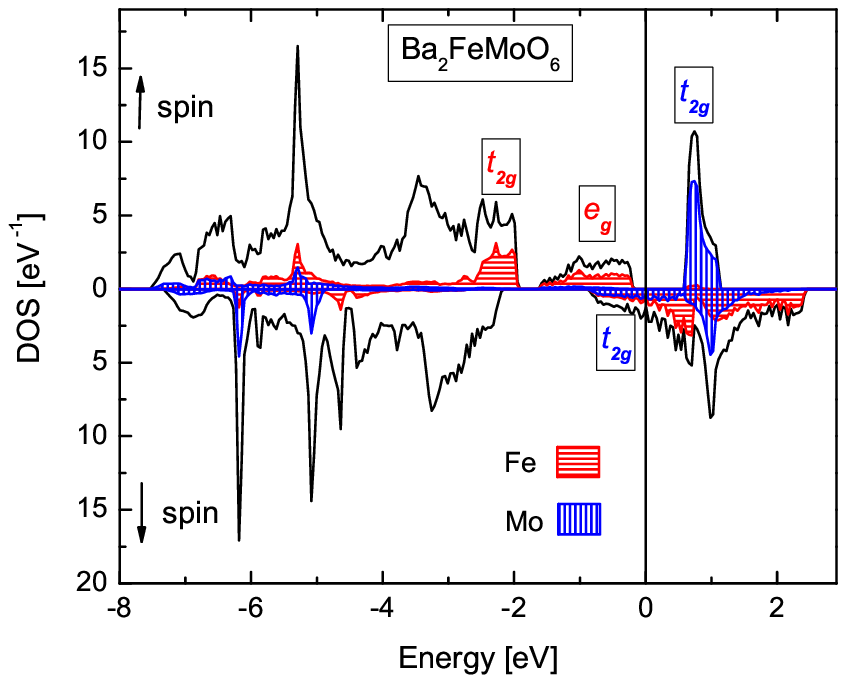,width=8cm}
 \epsfig{file=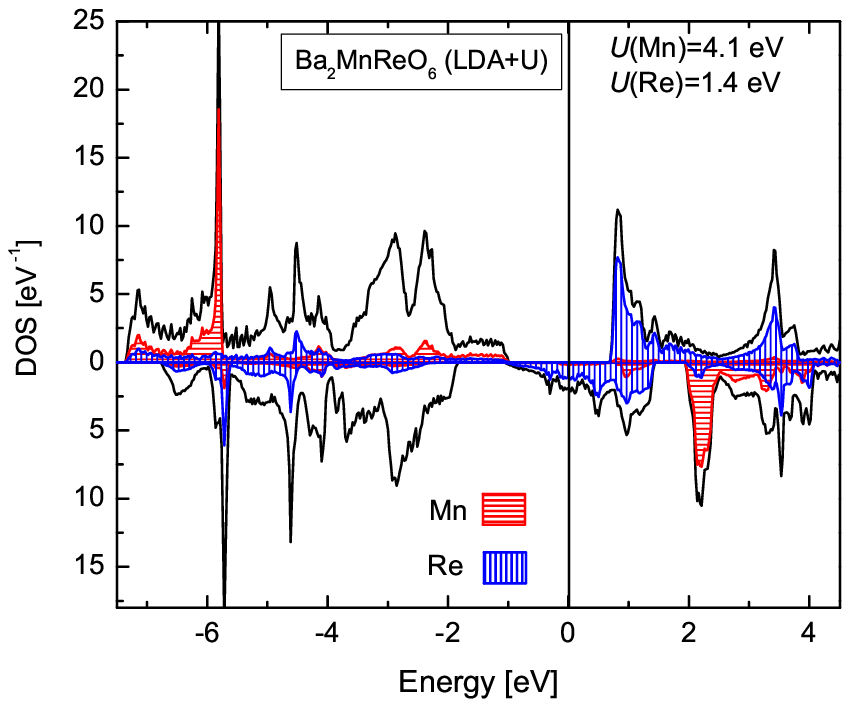,width=8cm}
 \caption{(Color online) Total (black)
and partial (colored) density of states of  Ba$_2$FeMoO$_6$ and
Ba$_2$MnReO$_6$. The number of conduction electrons from Fe and Mo
(Mn and Re) is $N_V$=6. The total magnetic moment is $M_{\rm
tot}=4.0$ $\mu_{\rm B}$ per f.u. The LDA+U method was employed in
the case of Ba$_2$MnReO$_6$.} \label{Dos-N14b}
\end{center}
\end{figure}

The double-perovskite series for $\rm B'$ = Fe and $N_V$ = 7, we
continue with Sr$_2$FeTcO$_6$ and Sr$_2$FeReO$_6$ for which the
densities of states are shown in Fig. \ref{Dos-N15a}. Both are
HMF. For Sr$_2$FeTcO$_6$ the Curie temperature we calculate to be
about 160 K, which is most likely underestimated. Sr$_2$FeReO$_6$
possesses a total magnetic moment of $M_{\rm{tot}}$ = 3 $\mu_{\rm
B}$, which is the result of a ferrimagnetic moment arrangement
comparable to the $N_V$ = 4 to 6 cases in Sr$_2$Cr$\rm B''O_6$
series. The DOS for Sr$_2$FeReO$_6$ shown in Fig. \ref{Dos-N15a}
is nearly identical with the previously published results by Fang
\textit{et al.}\cite{fang} Our calculation for the Curie
temperature gives $T_C$ = 384 K in good agreement with the
measured value of 400 K. We remark that the calculations were done
using an idealized cubic structure and the simple LSDA.

The case of $N_V$ = 8, Sr$_2$FeRuO$_6$, is again found to be
unstable as a ferromagnet with an unphysical spin-wave spectrum
(SWS). This means a branch of the SWS becomes negative indicating
that the assumed state is not the ground state. A search for the
correct state was carried out via the non-collinear states formed
by spin-spirals which lead to a total-energy minimum at
${\bf{K}}=(\frac {1}{2}, \frac{1}{2},\frac{1}{2})$. This is again
the AFII structure; it is metallic. As in the previous case for
Sr$_2$CrIrO$_6$, the metallic antiferromagnetic solution found
here has to be questioned. Thus, taking into account electron
correlations again by the LDA+U method we find Sr$_2$FeRuO$_6$ to
be a Mott-Hubbart insulator (MHI) with a small gap at the Fermi
energy, see Fig. \ref{Dos-N16}. Here the two Fe-peaks below -6 eV
are the filled $t_{2g}$- and $e_g$-states pushed to this low
energy by the Hubbard $U$. In contrast to the observations of Fang
\textit{et al.},\cite{fang} we find the magnetic moment of Ru to
be stable and arranged orthogonal to the Fe moments. However, an
attempt to determine the N\'eel temperature failed because the SWS
dipped into unphysical negative values again. Thus the MHI as
determined here still has to be questioned.

\begin{figure}[!]
\begin{center}
\epsfig{file=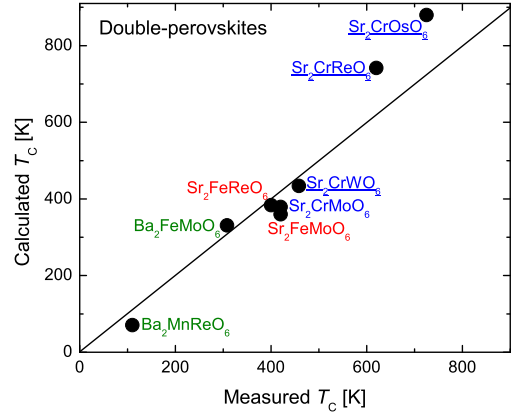,width=8cm} \caption{(Color online)
Calculated versus measured Curie temperatures. Underlined: GGA was
used; for Ba$_2$MnReO$_6$ LDA+U was used. } \label{Calc-mes-Tc}
\end{center}
\end{figure}

We finish with two double-perovskites having $N{_V}=6$:
Ba$_2$FeMoO$_6$ and Ba$_2$MnReO$_6$. Although somewhat outside the
series discussed above, we find them of importance for supporting
the predictive nature of our calculations.

Fig. \ref{Dos-N14b} shows that Ba$_2$FeMoO$_6$ is a HMF. Because
of the relatively large induced oxygen moments (see Table
\ref{t1}) we use six exchange functions and calculate the Curie
temperature to be $T_C$ = 331 K, which is to be compared with the
measured value of 308 K. In contrast to this rather
straight-forward LSDA-calculation is the case of Ba$_2$MnReO$_6$.

In both the LSDA and the GGA the electronic structure is
calculated to be a metallic ferrimagnet with spin-up
Mn-$t_{2g}$-states and spin-down Re-$e_g$-states at the Fermi
energy. This is in stark contrast to measurements by Popov
\textit{et al.} \cite{popov} who found this compound to be of high
resistivity. We therefore applied the LDA+U method and obtain a
HMF. The DOS is shown in the lower portion of Fig. \ref{Dos-N14b}.
Comparing with the case of Ba$_2$FeMoO$_6$ in the upper portion of
the Fig. \ref{Dos-N14b} we see that all spin-up Mn states are well
below $E_F$ and the states at $E_F$ are almost purely spin-down Re
states. The calculated Curie temperature is 71 K which is in fair
agreement with the measured value of 110 K.

In Fig. \ref{Calc-mes-Tc} we finally compare all measured and
calculated Curie temperatures.

\section{Summary\label{p4}}

\begin{figure}[!]
\begin{center}
 \epsfig{file=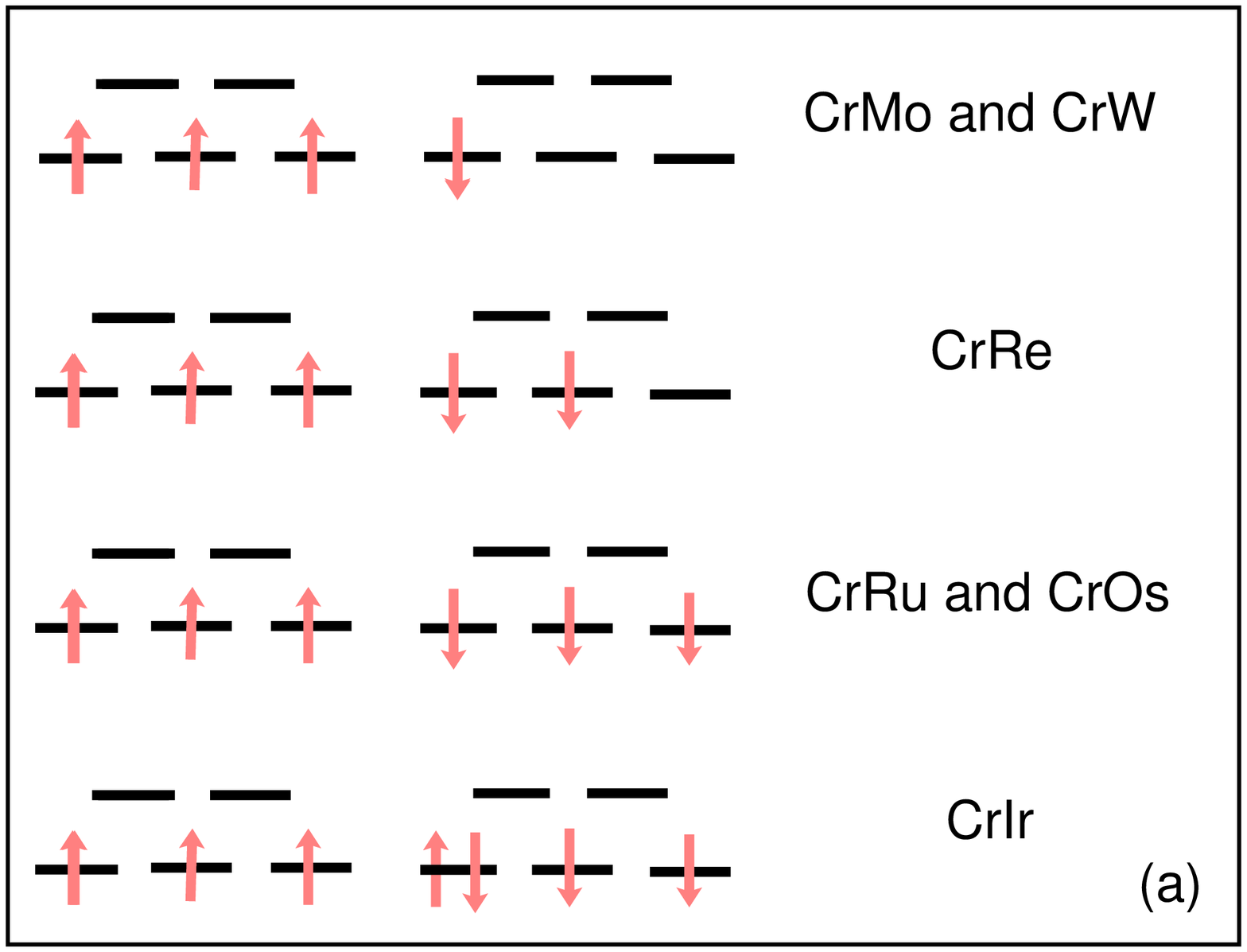,width=6cm}
 \epsfig{file=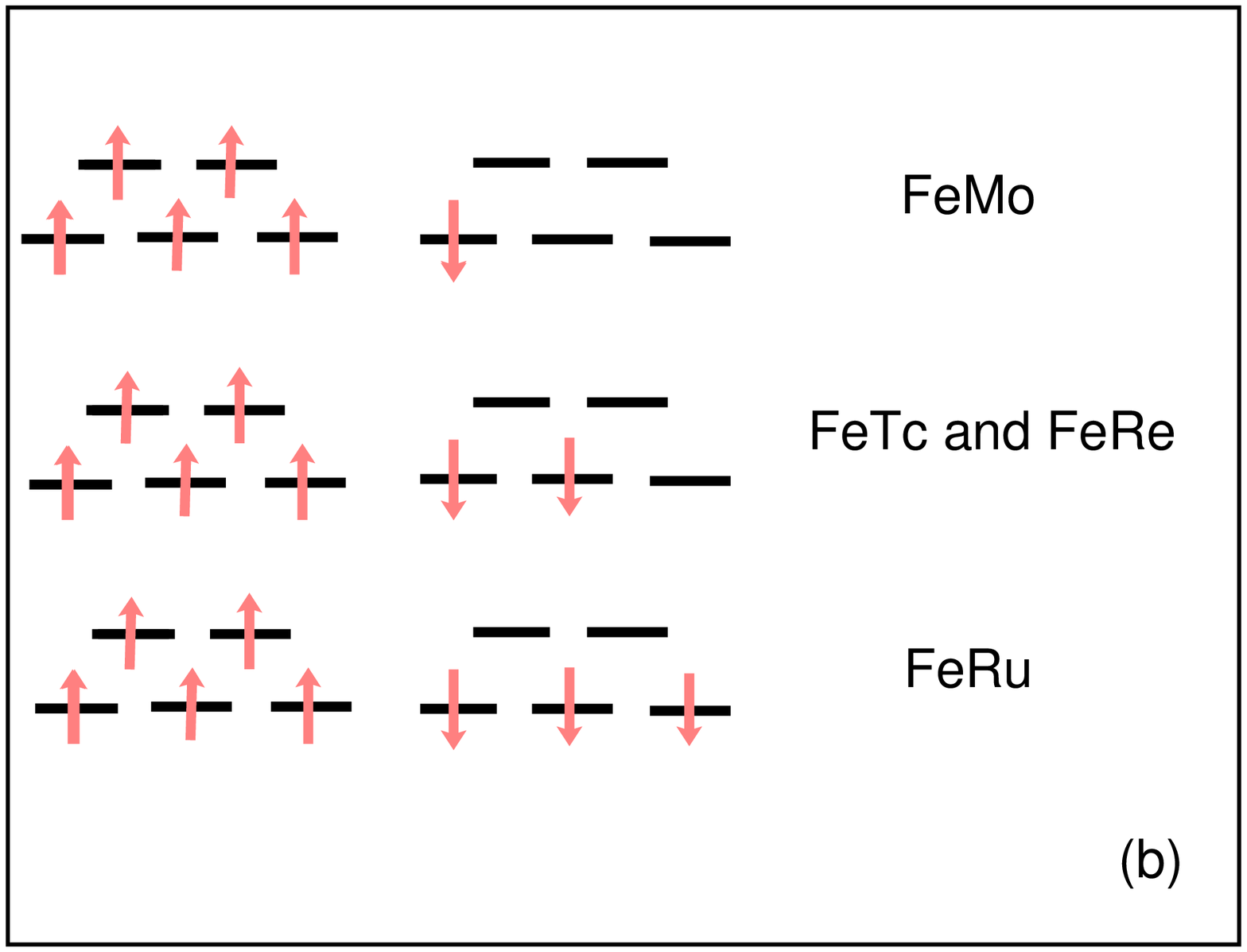,width=6cm}
 \caption{(Color online) Schematic level diagram for (a) the series
 Sr$_{2}\rm Cr\rm B''\rm O_6$ and (b) Sr$_{2}\rm Fe\rm B''\rm O_6$.
 Three lines symbolize the $t_{2g}$-states and two the $e_g$-states.
 The number of valence electrons, $N_V$, is given by the sum of the arrows,
 which (by our definition used in Figs. \ref{Dos-N12} to \ref{Dos-N14b}) count the
 up- and down-spin occupations.} \label{summ}
\end{center}
\end{figure}

An overall view of the electronic and magnetic structure of the
two series considered here is easily obtained by using the
simplified level scheme graphed in Fig. \ref{summ}, which takes
into account the crystal-field splitting of an octahedral
environment, viz. $t_{2g}$ and $e_g$.

Since the two transition metals are coupled antiferromagnetically
the double-perovskites can be half metallic ferrimagnets (HMF),
antiferromagnets or magnetic insulators (MIN), depending on the
valence electron count. The single exception is Sr$_{2}\rm Cr\rm
Ir\rm O_6$, where the magnetic interaction is more complex.
Comparing with Figs. \ref{Dos-N12} and \ref{Dos-N13}, we see that
 Fig. \ref{summ} (a) captures the physics well. When the cations
both have a spin configuration of $d^3$ the compound becomes a
MIN, as is the case for Sr$_{2}\rm Cr\rm Os\rm O_6$ and Sr$_{2}\rm
Cr\rm Ru\rm O_6$ shown in Fig. \ref{Dos-N14a}. The series
represented by Fig. \ref{summ} (b) is terminated by the
Mott-Hubbard insulator Sr$_{2}\rm Fe\rm Ru\rm O_6$ where the
cations have the spin configuration $d^5$ and $d^3$. The two
double-perovskites Ba$_{2}\rm Fe\rm Mo\rm O_6$ and Ba$_{2}\rm
Mn\rm Re\rm O_6$ shown in Fig. \ref{Dos-N14b} are represented in
Fig. \ref{summ} (b) by the FeMo entry.

Our \textit{ab initio} calculations of the Curie temperatures of
the double-perovskites reproduced the trend shown in Fig.
\ref{mes-Tc} even though we simplified the crystal structure by
using cubic cells. The spherical approximation together with the
use of spin-spirals proved to be quite reliable here. A glance at
Fig. \ref{Calc-mes-Tc} reveals deviations especially in the
calculations where the GGA had to be used. It is emphasized,
however, that the neglected structural distortions and anti-site
disorder can influence the measured magnetic properties so that a
comparison with the idealized calculations may only result in
qualitative understanding.

Finally, the large Curie temperatures, especially for Sr$_{2}\rm
Cr\rm Re\rm O_6$ and Sr$_{2}\rm Cr\rm Os\rm O_6$, require
remarkably large ferromagnetic exchange interactions between the
Cr cations. This issue was raised by Sarma \textit{et al.}
\cite{sarma} and in large detail by Fang \textit{et al.}
\cite{fang} who invoked a mechanism that is comparable to double
exchange (DE).\cite{anderson1,degennes} This mechanism involves
charge carriers and is thus operative for the half metallic
ferrimagnets. For the magnetic insulators Sr$_{2}\rm Cr\rm Os\rm
O_6$ and Sr$_{2}\rm Cr\rm Ru\rm O_6$ and the Mott-Hubbard
insulator Sr$_{2}\rm Fe\rm Ru\rm O_6$ we resort to the
antiferromagnetic superexchange (SE),\cite{anderson2} which
through the $4d$- and $5d$-cations results in ferromagnetic
exchange between the $3d$-cations. The interaction path that is
typical for both DE and SE possesses a distinctly visible
fingerprint in the Figs. \ref{Dos-N12} to \ref{Dos-N14b} where the
oxygen-transition-metal hybridization is clearly visible at an
energy of about -7 eV to -6 eV.

\begin{acknowledgments}
J.K. is grateful to L.M. Sandratskii for letting him use his LDA+U
code. T.K.M. and M.G. thank the National Science Foundation for
financial support through NSF-DMR-0233697 and NSF-DMR-0541911.
C.F. and J.K. acknowledge the financial support by the Deutsche
Forschungsgemeinschaft in research unit FG~559.
\end{acknowledgments}

\bibliography{kuebler}

\begin{thebibliography}{18}
\expandafter\ifx\csname natexlab\endcsname\relax\def\natexlab#1{#1}\fi
\expandafter\ifx\csname bibnamefont\endcsname\relax
  \def\bibnamefont#1{#1}\fi
\expandafter\ifx\csname bibfnamefont\endcsname\relax
  \def\bibfnamefont#1{#1}\fi
\expandafter\ifx\csname citenamefont\endcsname\relax
  \def\citenamefont#1{#1}\fi
\expandafter\ifx\csname url\endcsname\relax
  \def\url#1{\texttt{#1}}\fi
\expandafter\ifx\csname urlprefix\endcsname\relax\def\urlprefix{URL }\fi
\providecommand{\bibinfo}[2]{#2}
\providecommand{\eprint}[2][]{\url{#2}}

\bibitem[{\citenamefont{Kobayashi et al.}(1998)\citenamefont{Kobayashi, Kimura, Sawada, Terakura,
Tokura}}]{terakura}
\bibinfo{author}{\bibfnamefont{K.-I. }\bibnamefont{Kobayashi}}, %
 \bibinfo{author}{\bibfnamefont{ T. }\bibnamefont{Kimura}}, %
 \bibinfo{author}{\bibfnamefont{ H. }\bibnamefont{Sawada}}, %
 \bibinfo{author}{\bibfnamefont{ K. }\bibnamefont{Terakura}}, %
 \bibnamefont{ and} 
 \bibinfo{author}{\bibfnamefont{ Y. }\bibnamefont{Tokura}},
 \bibinfo{journal}{Nature} \textbf{\bibinfo{volume}{395}},
  \bibinfo{pages}{677} (\bibinfo{year}{1998}).



\bibitem[{\citenamefont{Okimoto et al.}(1983)}]{okimoto}
\bibinfo{author}{\bibfnamefont{Y.}~\bibnamefont{Okimoto}},
  \bibinfo{author}{\bibfnamefont{ T.} \bibnamefont{Katsufuji}},
    \bibinfo{author}{\bibfnamefont{ T.} \bibnamefont{Ishikawa}},
      \bibinfo{author}{\bibfnamefont{ A.} \bibnamefont{Urushibara}},
        \bibinfo{author}{\bibfnamefont{ T.} \bibnamefont{Arima}}
  , \bibnamefont{ and}
  \bibinfo{author}{\bibfnamefont{ Y.} \bibnamefont{Tokura}},
   \bibinfo{journal}{Phys. Rev. Lett.}
  \textbf{\bibinfo{volume}{75}}, \bibinfo{pages}{109} (\bibinfo{year}{1995}).

\bibitem[{\citenamefont{Park et al.}(1983)}]{park}
\bibinfo{author}{\bibfnamefont{J.~H.}~\bibnamefont{Park}},
  \bibinfo{author}{\bibfnamefont{ E.} \bibnamefont{Vescovo}},
    \bibinfo{author}{\bibfnamefont{ H.~J.} \bibnamefont{Kim}},
      \bibinfo{author}{\bibfnamefont{ C.} \bibnamefont{Kwon}},
        \bibinfo{author}{\bibfnamefont{ R.} \bibnamefont{Ramesh}}
  , \bibnamefont{ and}
  \bibinfo{author}{\bibfnamefont{ T.} \bibnamefont{Venkatesan}},
   \bibinfo{journal}{Nature}
  \textbf{\bibinfo{volume}{392}}, \bibinfo{pages}{794} (\bibinfo{year}{1998}).
  
\bibitem[{\citenamefont{de Groot et al.}(1983)}]{degroot}
\bibinfo{author}{\bibfnamefont{R.~A.}~\bibnamefont{de Groot}},
  \bibinfo{author}{\bibfnamefont{ F.~M.} \bibnamefont{Mueller}},
    \bibinfo{author}{\bibfnamefont{ P.~G.} \bibnamefont{van Engen}}
  , \bibnamefont{ and}
  \bibinfo{author}{\bibfnamefont{ K.~H.~J.} \bibnamefont{Buschow}},
   \bibinfo{journal}{Phas. Rev. Lett.}
  \textbf{\bibinfo{volume}{50}}, \bibinfo{pages}{2024} (\bibinfo{year}{1983}).  
  
  
  \bibitem[{\citenamefont{K\"ubler and Williams}(1983)}]{kubwill}
\bibinfo{author}{\bibfnamefont{J.}~\bibnamefont{K\"ubler}},
  \bibinfo{author}{\bibfnamefont{ A.~R.} \bibnamefont{Williams}}, \bibnamefont{ and}
  \bibinfo{author}{\bibfnamefont{ C.~B.} \bibnamefont{Sommers}},
   \bibinfo{journal}{Phys. Rev. B}
  \textbf{\bibinfo{volume}{28}}, \bibinfo{pages}{1745} (\bibinfo{year}{1983}).

\bibitem[{\citenamefont{Serrate et~al.}(2007)\citenamefont{Serrate, De Teesa, Ibarra}}]{serrate}
\bibinfo{author}{\bibfnamefont{D. } \bibnamefont{Serrate}},
  \bibinfo{author}{\bibfnamefont{ J.~M. } \bibnamefont{De Teresa}},
  \bibnamefont{ and} \bibinfo{author}{\bibfnamefont{ M.~R. }~\bibnamefont{Ibarra}},
  \bibinfo{journal}{J. Phys. CM: Condens. Matter} \textbf{\bibinfo{volume}{19}},
  \bibinfo{pages}{023201} (\bibinfo{year}{2007}).

\bibitem[{\citenamefont{K\"ubler et~al.}(2007)\citenamefont{K\"ubler, Fecher, Felser}}]{kublerxx}
\bibinfo{author}{\bibfnamefont{J.}~\bibnamefont{K\"ubler}},
  \bibinfo{author}{\bibfnamefont{ G.~H. } \bibnamefont{Fecher}},
  \bibnamefont{ and} \bibinfo{author}{\bibfnamefont{ C.}~\bibnamefont{Felser}},
  \bibinfo{journal}{Phys. Rev. B} \textbf{\bibinfo{volume}{76}},
  \bibinfo{pages}{024414} (\bibinfo{year}{2007}).


\bibitem[{\citenamefont{Kohn and Sham}(1965)}]{kohn}
\bibinfo{author}{\bibfnamefont{W.}~\bibnamefont{Kohn}} \bibnamefont{ and}
  \bibinfo{author}{\bibfnamefont{ L.~J.} \bibnamefont{Sham}},
  \bibinfo{journal}{Phys. Rev.} \textbf{\bibinfo{volume}{140}},
  \bibinfo{pages}{A1133} (\bibinfo{year}{1965}).

\bibitem[{\citenamefont{von Barth and Hedin}(1972)}]{vonbarth}
\bibinfo{author}{\bibfnamefont{U.}~\bibnamefont{von Barth}} \bibnamefont{ and}
  \bibinfo{author}{\bibfnamefont{ L.}~\bibnamefont{Hedin}}, \bibinfo{journal}{J.
  Phys. C: Sol. State Phys.} \textbf{\bibinfo{volume}{5}},
  \bibinfo{pages}{1629} (\bibinfo{year}{1972}).


\bibitem[{\citenamefont{Williams et~al.}(1979)\citenamefont{Williams, K\"ubler,
  and Gelatt}}]{williams}
\bibinfo{author}{\bibfnamefont{A.~R. } \bibnamefont{Williams}},
  \bibinfo{author}{\bibfnamefont{ J. } \bibnamefont{K\"ubler}}, \bibnamefont{ and}
  \bibinfo{author}{\bibfnamefont{ C.~D.} \bibnamefont{Gelatt}},
  \bibinfo{journal}{Phys. Rev. B} \textbf{\bibinfo{volume}{19}},
  \bibinfo{pages}{6094} (\bibinfo{year}{1979}).


\bibitem[{\citenamefont{Perdew et~al.}(1996)\citenamefont{Perdew, Burke, and
  Ernzerhof}}]{perdew}
\bibinfo{author}{\bibfnamefont{J.~P.} \bibnamefont{Perdew}},
  \bibinfo{author}{\bibfnamefont{ K.}~\bibnamefont{Burke}}, \bibnamefont{ and}
  \bibinfo{author}{\bibfnamefont{ M.}~\bibnamefont{Ernzerhof}},
  \bibinfo{journal}{Phys. Rev. Lett.} \textbf{\bibinfo{volume}{77}},
  \bibinfo{pages}{3865} (\bibinfo{year}{1996}).

\bibitem[{\citenamefont{Kn\"opfle et~al.}(2000)\citenamefont{Kn\"opfle,
  Sandratskii, and K\"ubler}}]{knopfle}
\bibinfo{author}{\bibfnamefont{K.}~\bibnamefont{Kn\"opfle}},
  \bibinfo{author}{\bibfnamefont{ L.~M. } \bibnamefont{Sandratskii}},
  \bibnamefont{ and} \bibinfo{author}{\bibfnamefont{ J.}~\bibnamefont{K\"ubler}},
  \bibinfo{journal}{Phys. Rev. B} \textbf{\bibinfo{volume}{62}},
  \bibinfo{pages}{5564} (\bibinfo{year}{2000}).


\bibitem[{\citenamefont{Popov et~al.}(2003)\citenamefont{Popov, Greenblatt, Croft}}]{popov}
\bibinfo{author}{\bibfnamefont{G.} \bibnamefont{Popov}},
  \bibinfo{author}{\bibfnamefont{ M.} \bibnamefont{Greenblatt}},
  \bibinfo{author}{\bibfnamefont{ M.}~\bibnamefont{Croft}},
  \bibinfo{journal}{Phys. Rev. B} \textbf{\bibinfo{volume}{67}},
  \bibinfo{pages}{024406} (\bibinfo{year}{2003}).

\bibitem[{\citenamefont{Anisimov et~al.}(1991)\citenamefont{Anisimov, Zaanen, Andersen}}]{anisimov}
\bibinfo{author}{\bibfnamefont{V.~I.}~\bibnamefont{Anisimov}},
  \bibinfo{author}{\bibfnamefont{ J. } \bibnamefont{Zaanen}},
  \bibnamefont{ and} \bibinfo{author}{\bibfnamefont{ O.~K.}~\bibnamefont{Andersen}},
  \bibinfo{journal}{Phys. Rev. B} \textbf{\bibinfo{volume}{44}},
  \bibinfo{pages}{943} (\bibinfo{year}{1991}).


\bibitem[{\citenamefont{Sandratskii}(2006)}]{sand}
\bibinfo{author}{\bibfnamefont{L.~M. } \bibnamefont{Sandratskii}},
  \bibinfo{journal}{unpublished} (\bibinfo{year}{2006}).

\bibitem[{\citenamefont{Fang et al.}(2001)\citenamefont{Fang, Terakura,
Kanamori}}]{fang}
\bibinfo{author}{\bibfnamefont{Z. }\bibnamefont{Fang}}, %
 \bibinfo{author}{\bibfnamefont{ K. }\bibnamefont{Terakura}}, %
 \bibnamefont{ and} 
 \bibinfo{author}{\bibfnamefont{ J. }\bibnamefont{Kanamori}},
 \bibinfo{journal}{Phys. Rev. B} \textbf{\bibinfo{volume}{63}},
  \bibinfo{pages}{180407(R)} (\bibinfo{year}{2001}).

\bibitem[{\citenamefont{Herring}(1966)}]{herring}
\bibinfo{author}{\bibfnamefont{C.}~\bibnamefont{Herring}}, in
  \emph{\bibinfo{booktitle}{Magnetism IV}}, edited by
  \bibinfo{editor}{\bibfnamefont{G.}~\bibnamefont{Rado}} \bibnamefont{and}
  \bibinfo{editor}{\bibfnamefont{H.}~\bibnamefont{Suhl}}
  (\bibinfo{publisher}{Academic Press}, \bibinfo{address}{New York},
  \bibinfo{year}{1966}).

\bibitem[{\citenamefont{Sandratskii}(1986)}]{sandratskii1}
\bibinfo{author}{\bibfnamefont{L.~M. } \bibnamefont{Sandratskii}},
  \bibinfo{journal}{J. Phys. F: Metal Phys.} \textbf{\bibinfo{volume}{16}},
  \bibinfo{pages}{L43} (\bibinfo{year}{1986}).


\bibitem[{\citenamefont{Sandratskii}(1998)}]{sandratskii}
\bibinfo{author}{\bibfnamefont{L.~M. } \bibnamefont{Sandratskii}},
  \bibinfo{journal}{Adv. Phys.} \textbf{\bibinfo{volume}{47}},
  \bibinfo{pages}{91} (\bibinfo{year}{1998}).


\bibitem[{\citenamefont{Popov et~al.}(2004)\citenamefont{Popov, Lobanov, Tsiper, Greenblatt,
Caspi, Borissov, Kiryukhin, Lynn}}]{popov1}
\bibinfo{author}{\bibfnamefont{G. }\bibnamefont{Popov}}, %
 \bibinfo{author}{\bibfnamefont{ V. }\bibnamefont{Lobanov}}, %
 \bibinfo{author}{\bibfnamefont{ E.V. }\bibnamefont{Tsiper}}, %
 \bibinfo{author}{\bibfnamefont{ M. }\bibnamefont{Greenblatt}}, %
 \bibinfo{author}{\bibfnamefont{ E.N. }\bibnamefont{Caspi}}, %
  \bibinfo{author}{\bibfnamefont{ A. }\bibnamefont{Borissov}}, %
   \bibinfo{author}{\bibfnamefont{ V. }\bibnamefont{Kiryukhin}}, %
  \bibnamefont{ and} 
 \bibinfo{author}{\bibfnamefont{ J.W. }\bibnamefont{Lynn}},
 \bibinfo{journal}{J. Phys.: Condens. Matter} \textbf{\bibinfo{volume}{16}},
  \bibinfo{pages}{135} (\bibinfo{year}{2004}).

\bibitem[{\citenamefont{K{\"u}bler}(2006)}]{kublerx}
\bibinfo{author}{\bibfnamefont{J.}~\bibnamefont{K{\"u}bler}},
  \bibinfo{journal}{J. Phys.: Condens. Matter} \textbf{\bibinfo{volume}{18}},
  \bibinfo{pages}{9795} (\bibinfo{year}{2006}).



\bibitem[{\citenamefont{Heine}(1980)}]{heine}
\bibinfo{author}{\bibfnamefont{V.}~\bibnamefont{Heine}}, in
  \emph{\bibinfo{booktitle}{Solid State Physics}}, edited by
  \bibinfo{editor}{\bibfnamefont{H.}~\bibnamefont{Ehrenreich}},
  \bibinfo{editor}{\bibfnamefont{F.}~\bibnamefont{Seitz}}, \bibnamefont{and}
  \bibinfo{editor}{\bibfnamefont{P.}~\bibnamefont{Turnbull}}
  (\bibinfo{publisher}{Aademic Press}, \bibinfo{address}{New York},
  \bibinfo{year}{1980}), vol.~\bibinfo{volume}{35}.

\bibitem[{\citenamefont{Mackintosh and Andersen}(1980)}]{mackintosh}
\bibinfo{author}{\bibfnamefont{A.~R. } \bibnamefont{Mackintosh}}
  \bibnamefont{and} \bibinfo{author}{\bibfnamefont{O.~K. }
  \bibnamefont{Andersen}}, in \emph{\bibinfo{booktitle}{Electrons at the Fermi
  surface}}, edited by \bibinfo{editor}{\bibfnamefont{S.}~\bibnamefont{M.}}
  (\bibinfo{publisher}{Cambridge University Press},
  \bibinfo{address}{Cambridge}, \bibinfo{year}{1980}).


\bibitem[{\citenamefont{Moriya}(1985)}]{moriya}
\bibinfo{author}{\bibfnamefont{T.}~\bibnamefont{Moriya}},
  \emph{\bibinfo{title}{Spin Fluctuations in Itinerant Electron Magnetism}}
  (\bibinfo{publisher}{Springer Verlag}, \bibinfo{address}{Berlin},
  \bibinfo{year}{1985}).


\bibitem[{\citenamefont{Arulraj et~al.}(2000)}]{arul}
\bibinfo{author}{\bibfnamefont{A.}~\bibnamefont{Arulraj}},
  \bibinfo{author}{\bibfnamefont{K. } \bibnamefont{Ramesha}},
  \bibinfo{author}{\bibfnamefont{J.}~\bibnamefont{Gopalakrishnan}},
  \bibnamefont{ and } 
  \bibinfo{author}{\bibfnamefont{C.N.N.}~\bibnamefont{Rao}},
  \bibinfo{journal}{J. Sol. State Chem.} \textbf{\bibinfo{volume}{155}},
  \bibinfo{pages}{233} (\bibinfo{year}{2000}).

\bibitem[{\citenamefont{Kato et~al.}(2007)}]{kato}
\bibinfo{author}{\bibfnamefont{H.}~\bibnamefont{Kato}},
  \bibinfo{author}{\bibfnamefont{T. } \bibnamefont{Okuda}},
  \bibinfo{author}{\bibfnamefont{Y.}~\bibnamefont{Okimoto}},
  \bibinfo{author}{\bibfnamefont{Y.}~\bibnamefont{Tomioka}},
  \bibinfo{author}{\bibfnamefont{K.}~\bibnamefont{Oikawa}},
  \bibinfo{author}{\bibfnamefont{T.}~\bibnamefont{Kamiyama}},
  \bibnamefont{ and } 
  \bibinfo{author}{\bibfnamefont{Y.}~\bibnamefont{Tokura}},
  \bibinfo{journal}{Phys. Rev. B} \textbf{\bibinfo{volume}{69}},
  \bibinfo{pages}{184412} (\bibinfo{year}{2004}).

\bibitem[{\citenamefont{Krockenberger et~al.}(2007)\citenamefont{Krockenberger, Mogare,
Reehuis, Tovar, Jansen, Vaitheeswaran, Kanchana, Bultmark, Delin, Wilhelm,
Rofalev, Winkler, Alff}}]{krocken}
\bibinfo{author}{\bibfnamefont{Y.}~\bibnamefont{Krockenberger}},
  \bibinfo{author}{\bibfnamefont{K. } \bibnamefont{Mogare}},
  \bibinfo{author}{\bibfnamefont{M.}~\bibnamefont{Reehuis}},
  \bibinfo{author}{\bibfnamefont{M.}~\bibnamefont{Tovar}},
  \bibinfo{author}{\bibfnamefont{M.}~\bibnamefont{Jansen}},
  \bibinfo{author}{\bibfnamefont{G.}~\bibnamefont{Vaitheeswaran}},
  \bibinfo{author}{\bibfnamefont{V.}~\bibnamefont{Kanchana}},
  \bibinfo{author}{\bibfnamefont{F.}~\bibnamefont{Bultmark}},
  \bibinfo{author}{\bibfnamefont{A.}~\bibnamefont{Delin}},
  \bibinfo{author}{\bibfnamefont{F.}~\bibnamefont{Wilhelm}},
  \bibinfo{author}{\bibfnamefont{A.}~\bibnamefont{Rogalev}},
  \bibinfo{author}{\bibfnamefont{A.}~\bibnamefont{Winkler}},
  \bibnamefont{ and } 
  \bibinfo{author}{\bibfnamefont{L.}~\bibnamefont{Alff}},
  \bibinfo{journal}{Phys. Rev. B} \textbf{\bibinfo{volume}{75}},
  \bibinfo{pages}{020404(R)} (\bibinfo{year}{2007}).

  



\bibitem[{\citenamefont{Lee and Pickett}(2007)\citenamefont{Lee and Pickett}}]{lee}
\bibinfo{author}{\bibfnamefont{K.-W. }\bibnamefont{Lee}} %
 \bibnamefont{and} 
 \bibinfo{author}{\bibfnamefont{ W.~E. }\bibnamefont{Pickett}},
 \bibinfo{journal}{Phys. Rev. B} \textbf{\bibinfo{volume}{77}},
  \bibinfo{pages}{115101} (\bibinfo{year}{2008}).

 
\bibitem[{\citenamefont{Michalik et al.}(2007)\citenamefont{Michalik et al.}}]{michalik}
\bibinfo{author}{\bibfnamefont{J.~M. }\bibnamefont{Michalik}}, 
\bibinfo{author}{\bibfnamefont{ J.~M. }\bibnamefont{De Teresa}},
\bibinfo{author}{\bibfnamefont{ C. }\bibnamefont{Ritter}},
\bibinfo{author}{\bibfnamefont{ J. }\bibnamefont{Blasco}},
\bibinfo{author}{\bibfnamefont{ D. }\bibnamefont{Serrate}},
\bibinfo{author}{\bibfnamefont{ M.~R. }\bibnamefont{Ibarra}}, 
\bibinfo{author}{\bibfnamefont{ C. }\bibnamefont{Kapusta}},  
\bibinfo{author}{\bibfnamefont{ J. }\bibnamefont{Freudenberger}},    
 \bibnamefont{ and} 
 \bibinfo{author}{\bibfnamefont{ N. }\bibnamefont{Kozlova}},
 \bibinfo{journal}{Europhys. Lett.} \textbf{\bibinfo{volume}{78}},
  \bibinfo{pages}{17006} (\bibinfo{year}{2007}).

\bibitem[{\citenamefont{De Teresa et al.}(2007)\citenamefont{De Teresa et al.}}]{teresa}
\bibinfo{author}{\bibfnamefont{J.~M. }\bibnamefont{De Teresa}}, 
\bibinfo{author}{\bibfnamefont{ J.~M. }\bibnamefont{Michalik}},
\bibinfo{author}{\bibfnamefont{ J. }\bibnamefont{Blasco}},
\bibinfo{author}{\bibfnamefont{ P.~A. }\bibnamefont{Algarabel}},
\bibinfo{author}{\bibfnamefont{ M.~R. }\bibnamefont{Ibarra}}, 
\bibinfo{author}{\bibfnamefont{ C. }\bibnamefont{Kapusta}},      
 \bibnamefont{ and} 
 \bibinfo{author}{\bibfnamefont{ U. }\bibnamefont{Zeitler}},
 \bibinfo{journal}{Appl. Phys. Lett.} \textbf{\bibinfo{volume}{90}},
  \bibinfo{pages}{252514} (\bibinfo{year}{2007}).

\bibitem[{\citenamefont{Vaitheeswaran et al.}(2007)\citenamefont{Vaitheeswaran et al.}}]{vaitheeswaran}
\bibinfo{author}{\bibfnamefont{G. }\bibnamefont{Vaitheeswaran}}, 
\bibinfo{author}{\bibfnamefont{ V. }\bibnamefont{Kanchana}},
 \bibnamefont{ and} 
 \bibinfo{author}{\bibfnamefont{ A. }\bibnamefont{Delin}},
 \bibinfo{journal}{Appl. Phys. Lett.} \textbf{\bibinfo{volume}{86}},
  \bibinfo{pages}{032513} (\bibinfo{year}{2005}).

\bibitem[{\citenamefont{Sarma et al.}(2000)\citenamefont{Sarma et al.}}]{sarma}
\bibinfo{author}{\bibfnamefont{D.~D. }\bibnamefont{Sarma}}, 
\bibinfo{author}{\bibfnamefont{ P. }\bibnamefont{Mahadevan}},
\bibinfo{author}{\bibfnamefont{ T.~S. }\bibnamefont{Dasgupta}},
\bibinfo{author}{\bibfnamefont{ S. }\bibnamefont{Ray}},       
 \bibnamefont{ and} 
 \bibinfo{author}{\bibfnamefont{ A. }\bibnamefont{Kumar}},
 \bibinfo{journal}{Phys. Rev. Lett.} \textbf{\bibinfo{volume}{85}},
  \bibinfo{pages}{2549} (\bibinfo{year}{2000}).

\bibitem[{\citenamefont{Anderson and Hasegawa}(1955)\citenamefont{Anderson and Hasegawa}}]{anderson1}
\bibinfo{author}{\bibfnamefont{P.~W. }\bibnamefont{Anderson}}, 
 \bibnamefont{ and} 
 \bibinfo{author}{\bibfnamefont{ H. }\bibnamefont{Hasegawa}},
 \bibinfo{journal}{Phys. Rev. } \textbf{\bibinfo{volume}{100}},
  \bibinfo{pages}{675} (\bibinfo{year}{1955}).
  
\bibitem[{\citenamefont{de Gennes}(1960)\citenamefont{de Gennes}}]{degennes}
\bibinfo{author}{\bibfnamefont{P.-G. }\bibnamefont{de Gennes}}, 
\bibinfo{journal}{Phys. Rev.} \textbf{\bibinfo{volume}{118}},
  \bibinfo{pages}{141} (\bibinfo{year}{1960}).
  
\bibitem[{\citenamefont{Anderson}(1960)\citenamefont{Anderson}}]{anderson2}
\bibinfo{author}{\bibfnamefont{P.~W. }\bibnamefont{Anderson}}, 
\bibinfo{journal}{Phys. Rev.} \textbf{\bibinfo{volume}{79}},
  \bibinfo{pages}{350} (\bibinfo{year}{1950}).  
\end{thebibliography}

\end{document}